\documentclass[a4paper,11pt]{article}
\usepackage{pos}

\usepackage[utf8]{inputenc}
\usepackage[T1]{fontenc}
\usepackage{graphicx}
\usepackage{enumitem}
\usepackage{natbib}
\usepackage{gensymb}
\usepackage{amsmath}
\usepackage{amssymb}
\usepackage{mathtools}
\usepackage{slashed}
\usepackage{aps_macros}

\title{Probing the properties of superheavy dark matter annihilating or decaying into neutrinos with ultra-high energy neutrino experiments}
 \ShortTitle{Probing the properties of superheavy dark matter with neutrino experiments}

\author*[a,b]{Claire Gu\'epin}
\author[c]{Roberto Aloisio}
\author[d,e]{Luis A. Anchordoqui}
\author[c]{Austin Cummings}
\author[b,f]{John F. Krizmanic}
\author[g]{Angela V. Olinto}
\author[h]{Mary Hall Reno}
\author[b]{Tonia M. Venters}

\affiliation[a]{Department of Astronomy, University of Maryland, College Park, MD, USA}
\affiliation[b]{NASA Goddard Space Flight Center, Greenbelt, MD, USA}
\affiliation[c]{Gran Sasso Science Institute, L'Aquila, Italy}
\affiliation[d]{Department of Physics and Astronomy, Lehman College, City University of New York,  NY NY 10468, USA}
\affiliation[e]{Department of Astrophysics, American Museum of Natural History, NY 10024, USA}
\affiliation[f]{Center for Space Science \& Technology, University of Maryland, Baltimore County, Baltimore, MD, USA}
\affiliation[g]{Department of Astronomy \& Astrophysics, KICP, EFI, The University of Chicago, Chicago, IL 60637, USA}
\affiliation[h]{Department of Physics and Astronomy, University of Iowa, Iowa City, IA, USA}

\emailAdd{cguepin@umd.edu}

\abstract{The evidence for dark matter particles, $\chi$, is compelling based on Galactic to cosmological scale observations. Thus far, the promising weakly interacting massive particle scenario have eluded detection, motivating alternative models of dark matter. We consider scenarios involving superheavy dark matter (SHDM) that potentially can decay or annihilate to neutrinos and antineutrinos. In the mass range $m_\chi=10^7-10^{15}\,{\rm GeV}$, we evaluate the sensitivities of future observatories POEMMA and GRAND for indirect dark matter detection via the measurement of neutrino-induced extensive air showers (EAS), compute the Auger and ANITA limits using their last up-to-date sensitivities, and compare them with IceCube limits. We also show that the uncertainties related to the dark matter distribution in the Galactic halo have a large impact on the neutrino flux. We show that a ground-based radio detector such as GRAND can achieve high sensitivities due to its large effective area and high duty cycle. Space-based Cherenkov detectors such as POEMMA that measure the EAS optical Cherenkov signal have the advantage of full-sky coverage and rapid slewing, enabling an optimized SHDM observation strategy focusing on the Galactic Center. We show that increasing the field of view of the Cherenkov detectors can significantly enhance the sensitivity. Moreover, POEMMA’s fluorescence observation mode that measures EAS above $20$\,EeV will achieve state-of-the-art sensitivity to SHDM properties at the highest mass scales.}

\FullConference{37$^{\rm{th}}$ International Cosmic Ray Conference (ICRC 2021)\\
		July 12th -- 23rd, 2021\\
		Online -- Berlin, Germany}

\begin{document}
\maketitle

\section{Introduction}

The existence of dark matter is established by a variety of astrophysical probes \citep{BHS05}, such as galaxy rotation curves \citep{Rubin85} or observations of large-scale structures \citep{Percival10}. Despite constituting $84\%$ of the matter density in the Universe, the nature of non-baryonic dark matter remains elusive, with a large number of candidates proposed, e.g. axions, neutralinos or superheavy dark matter \citep{Ellis00}. This diversity requires the development of various strategies for detection \citep{Feng:2010gw}, including collider experiments, direct and indirect detection experiments and astrophysical probes.

In this study, we focus on very-high- and ultra-high-energy neutrino searches as indirect probes for superheavy dark matter $\chi$ annihilation and decay. Two caveats of this analysis are to be mentioned. First, neutrino signals from dark matter decay are expected to be accompanied by electromagnetic signals, thus the region of the parameter space probed with neutrino detectors can also be tested by gamma-ray and cosmic-ray detectors \cite{Blanco:2018esa}. Second, we focus on annihilation for dark matter masses above the unitarity bound $m_\chi \sim 110\,{\rm TeV}$ \cite{Blum:2014dca}, remaining agnostic about the specific dark matter model enabling to evade this bound \cite[e.g.][]{DelleRose:2017vvz}.

High- and ultra-high-energy gamma-ray, cosmic-ray and neutrino detectors can provide strong constraints on the dark matter annihilation cross sections and decay widths, and various detectors and channels have been considered in previous work \citep[e.g.][]{IceCube18, Arguelles19}. In this study, we focus on the annihilation and decay channels $\chi \chi \rightarrow \nu \bar{\nu}$ and $\chi \rightarrow \nu \bar{\nu}$. The detectors IceCube, Auger and ANITA provide currently the most constraining limits from neutrino detection in the mass range $m_\chi = 10^7 - 10^{15}\,{\rm eV}$. We compute the sensitivities of POEMMA \citep{POEMMA_JCAP} and GRAND \citep{GRAND_WP} and update the limits of Auger \citep{Aab:2019auo}, ANITA \citep{ANITAIV19} with the most up-to-date sensitivities, compared with existing limits from IceCube \citep{Gaisser:2014foa, IceCube18}. We study several observation strategies for POEMMA, and highlight the importance of uncertainties related to the dark matter distribution in the Galactic halo.

\section{Production of neutrinos from superheavy dark matter annihilation or decay}\label{sec:model}

The neutrino flux from dark matter annihilation or decay depends on the spectrum of secondary products ${\rm d}N/{\rm d}E$, the rate $\Gamma$, and the integral along the line of sight $x$ of the dark matter density $\rho_\chi$
\begin{equation}\label{eq:intensity}
    \frac{{\rm d}\Phi}{{\rm d}\Omega {\rm d}E} \equiv \frac{\Gamma}{4 \pi m_\chi^a} \frac{{\rm d}N}{{\rm d}E} \, \int_{\rm l.o.s.} {\rm d}x \, \rho_\chi^a (x) \,.
\end{equation}
We focus on the dominant contribution from dark matter in the Galactic halo, and consider both the Burkert profile \cite{Nesti13} and the generalized NFW profile \cite{Benito20}, with a fraction $f_\chi = \rho_\chi/\rho_{\rm DM}$ of dark matter being superheavy. For the Burkert profile, $\rho_\chi (r) = f_\chi \, \rho_H \left( 1+r/R_H \right)^{-1} \left[ 1+\left(r/R_H\right)^2 \right]^{-1}$, with a central dark matter density $\rho_H \simeq 4 \times 10^7 M_\odot \, {\rm kpc}^{-3}$ and a core radius $R_H \simeq 9 \, {\rm kpc}$. The generalized NFW profile is characterized by $\rho_\chi (r) = f_\chi \, \rho_s \left( r/R_s \right)^{-\gamma} \left( 1 + r/R_s \right)^{-3+\gamma} \, , $
with $\rho_s = \rho_0 \left( R_0/R_s \right)^{\gamma} \left( 1 + R_0/R_s \right)^{3-\gamma}$, the local density $\rho_0 = 0.6\,{\rm GeV\,cm}^{-3}$, the slope $\gamma = 0.4$ and the scale radius $R_s = 8 \times 10^1 \,{\rm kpc}$, for the best fit parameters from \cite{Benito20}. The distance between the Sun and the Galactic Center is $R_0 = 8.178\,{\rm kpc}$ \cite{Gravity19}. For the channels considered in this study, $\chi \chi \rightarrow \nu \bar{\nu}$ for annihilation (where $\chi$ is its own antiparticle) and $\chi \rightarrow \nu \bar{\nu}$ for decay, the spectra of secondary neutrinos peaks at $E_\nu = m_\chi$ for annihilation and $E_\nu = m_\chi/2$ for decay \citep{Bauer20}, and is well described by a delta-function approximation. For annihilation, the rate is $\Gamma = \langle \sigma v \rangle /2$ with $\langle \sigma v \rangle \equiv \sum_i\sigma(\chi\chi\to \nu_\alpha \bar{\nu}_\alpha)$ the thermally averaged cross section for $\alpha = e,\ \mu,\ \tau$, with equal cross sections for the three neutrino flavors, and $a=2$. For decay, the rate $\Gamma = \Gamma_{\chi\rightarrow\nu\bar{\nu}}$ is the decay width, and $a=1$. In what follows, we refer to neutrinos and antineutrinos together as ``neutrinos''.

The number of detectable tau neutrinos $N_{\nu_\tau}$ for a given neutrino detector is calculated by combining equation~\ref{eq:intensity} with the time averaged effective area $\langle A_{\rm eff}(\Omega, E_\nu)\rangle$ and observation time $T_{\rm obs}$ of the detector. Considering delta-function approximations for the spectrum of secondary neutrinos, 
\begin{eqnarray}\label{eq:num_tau}
    N_{\nu_\tau} (E_\nu) &=& \frac{1}{\mathcal{N}_\nu} \frac{1}{4\pi}\frac{\mathcal{R}}{E_\nu^a} \int {\rm d} \Omega \int_{\rm l.o.s.} {\rm d}x \, \rho^a_\chi (x) \, 
    \langle A_{\rm eff}(\Omega,E_\nu)\rangle\, T_{\rm obs}\,.
\end{eqnarray}
For annihilation, $\mathcal{R} = \langle \sigma v \rangle$ and $a=2$, and for decay, $\mathcal{R} = \Gamma_{\chi\rightarrow\nu\bar{\nu}}$ and $a=1$. To estimate the sensitivities or limits of very-high- to ultra-high neutrino detectors to the dark matter annihilation cross section or decay width, we set $N_{\nu} = 2.44$, which corresponds to the $90\%$ C.L. limit with negligible background. Depending on the detection techniques considered, $N_{\nu} = 2.44$ corresponds to the number of tau neutrinos (POEMMA Cherenkov, GRAND) or to the total number of neutrinos $\sum_\alpha N_{\nu_\alpha} = 2.44$ (POEMMA fluorescence, Auger, ANITA IV). In section~\ref{sec:detection}, we describe the properties of the detectors considered in this work to constrain the dark matter properties.

\section{Observation with very-high and ultra-high neutrinos detectors}\label{sec:detection}

In this work, we calculate the sensitivities of the POEMMA and GRAND detectors to dark matter properties, update the limits from Auger and ANITA using their most up-to-date sensitivities, and compare these sensitivities and limits with the limits from IceCube. POEMMA (Probe of Extreme Multi-Messenger Astrophysics) and GRAND (Giant Radio Array for Neutrino Detection) are two future observatories aiming at detecting very to ultra-high energy neutrinos ($\gtrsim 10^7\,{\rm GeV}$). POEMMA will operate from space with two satellites equipped with Cherenkov (POEMMA C) and fluorescence (POEMMA F) detectors. GRAND will be deployed on the ground, with arrays of $10$k (GRAND10k) to $200$k (GRAND200k) radio antennas operating in the $50-200\,{\rm MHz}$ range. Cherenkov and radio signals are generated by extensive air showers from up-going $\tau$-lepton decays, and fluorescence signals are produced by interactions of neutrinos from all flavors in the atmosphere. In the Cherenkov observation mode, POEMMA can access the full-sky after a few precession periods and can adopt specific observation strategies due to the slewing capability of its detectors. In the fluorescence observation mode, POEMMA will achieve a ground-breaking sensitivity to neutrinos in the range $\sim 10^{11}-10^{15}\,{\rm GeV}$. GRAND will achieve a competitive diffuse sensitivity in the range $\sim 10^8-10^{11}\,{\rm GeV}$, together with a high duty-cycle in radio quiet areas. The locations of the antenna arrays will determine the sky coverage of the observatory.

For POEMMA's Cherenkov observation mode, we consider the effective area from \citep{Motloch:2013kva, Reno:2019jtr, Venters19}, $A_{\rm eff}(\Omega, E_\nu,t) = \int {\rm d}P_{\rm obs}(\Omega,E_\nu,s,t) A_{\rm Ch}(s)$. Possible backgrounds are discussed in \cite{Venters19}. The effective area depends on $A_{\rm Ch}(s)$, the area of the extensive air shower's Cherenkov cone subtended on the ground normal to the shower axis, where $s$ is the path length of the tau-lepton before its decay, and the differential observation probability ${\rm d}P_{\rm obs}$, which depends on the probability of the tau-lepton to exit the Earth given an incident energy and angle of the tau neutrino, on the tau-lepton decay probability as a function of $s$ and on the detection probability given the shower energy, altitude and angle.

\begin{figure}[ht]
    \centering
    \includegraphics[width=0.49\textwidth]{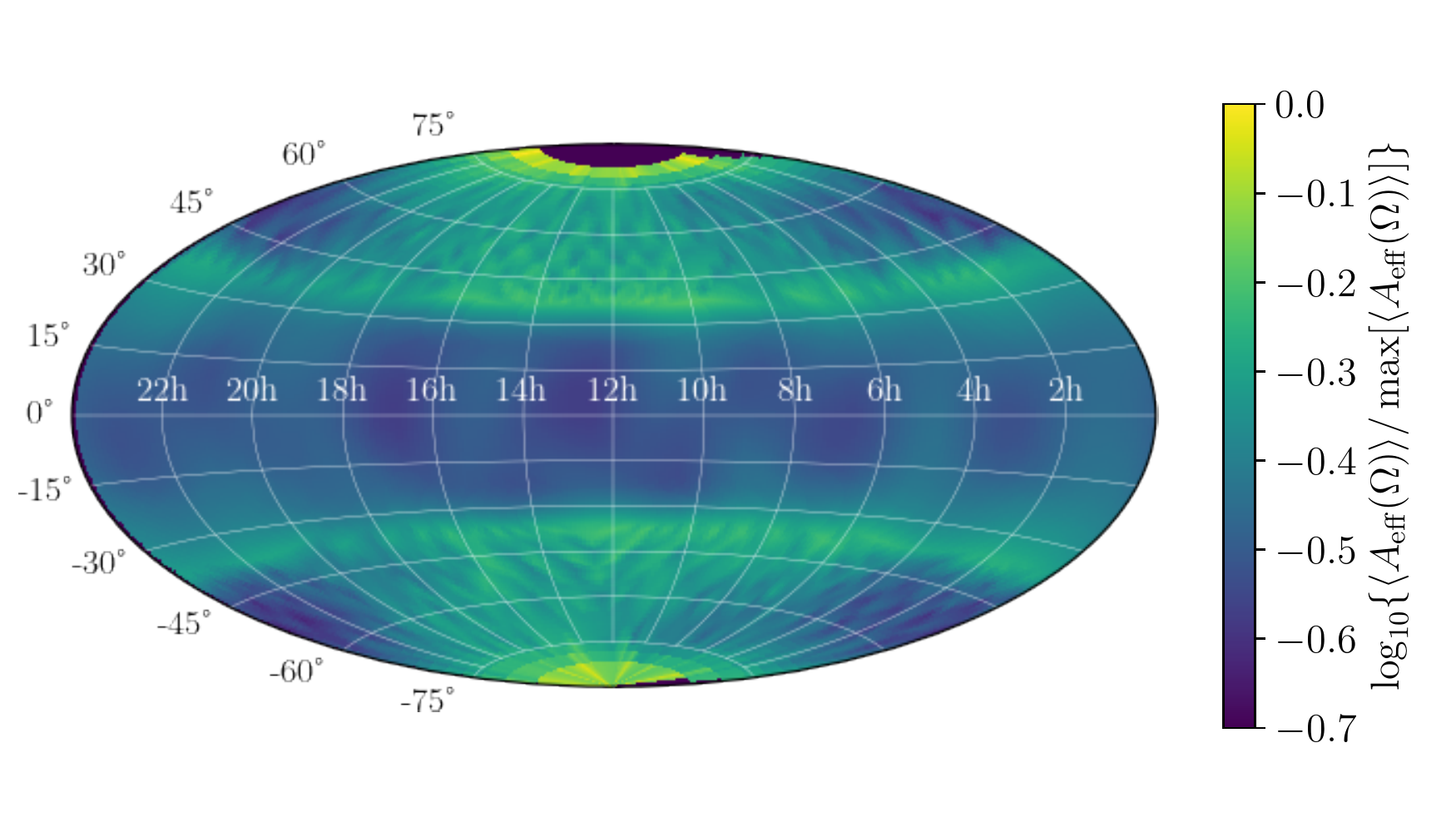}
    \includegraphics[width=0.49\textwidth]{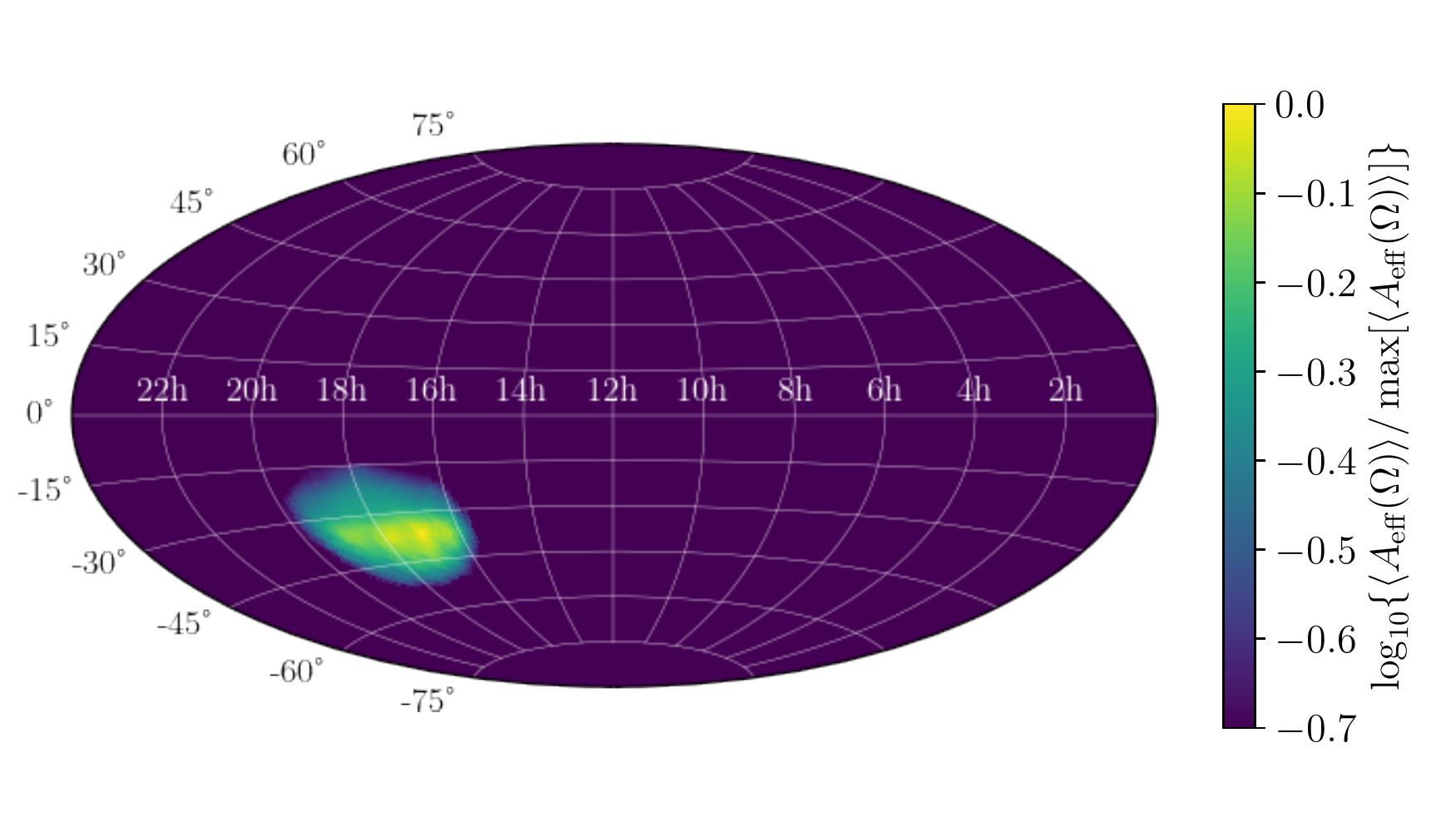}
    \caption{Normalized time averaged effective areas in logarithmic scale, for the standard observation mode (left) and the Galactic center observation mode (right), for $E_\nu = 10^{8.5}\,{\rm GeV}$.}\label{fig:Exp_comp}
\end{figure}

As the dark matter indirect detection is enhanced in the Galactic center direction, we develop a specific observation strategy for POEMMA's Cherenkov observation mode, by selecting the observable portion of the sky closest to the Galactic center. We combine sky coverage calculations accounting for the detector field of view and orientation \citep{Guepin19} with calculations of the best achievable differential exposure for every direction of the sky \citep{Venters19}. We name this strategy POEMMA Cherenkov Galactic center observation mode (POEMMA C, GC), while the standard observation mode (POEMMA C, std) refers to the observation strategy leading to a full-sky coverage. The time averaged effective areas for these two observation strategies are illustrated in Fig.~\ref{fig:Exp_comp}.

For GRAND10k and GRAND200k, we consider the GRAND differential effective areas for eight energy bins between $10^8\,{\rm GeV}$ and $10^{11.5}\,{\rm GeV}$, derived for an antenna array located at $43\degree$ latitude North (Olivier Martineau, private communication).

When the differential exposure of the detector is not directly available in the literature, namely for the fluorescence observation mode of POEMMA, Auger and ANITA-IV, we use the sensitivities of these detectors $F_\nu = E_\nu \, {\rm d} N_\nu / ({\rm d}E_\nu  \, {\rm d}A \, {\rm d} \Omega \, {\rm d}t)$ to compute the total exposure for one neutrino flavor $\mathcal{E} = 2.44 \, \mathcal{N}_\nu / [\ln(10) \, T_{\rm obs} \, 4\pi \, F_\nu ]$, with $\mathcal{N}_\nu = 3$ the number of neutrino flavors. The exposure is then combined with the sky coverage of the detector. For the fluorescence observation mode of POEMMA, the sensitivity \cite{Anchordoqui19, POEMMA_JCAP} uses two different neutrino cross sections, labeled GQRS \cite{GQRS98} and BDH \cite{BDH14}, and a uniform differential exposure over the entire sky is considered. The total exposure of Auger \citep{Aab15} is multiplied by a factor $1.5$ to account for the increase of exposure with time, and combined with the average neutrino exposure per day as a function of declination \citep{Aab19}. Finally, the sensitivity of ANITA-IV \cite{ANITAIV19} is combined with the ANITA-III effective area as a function of declination \cite{ANITA20}.

\section{Sensitivities to dark matter properties}

\begin{figure}[ht]
    \centering
    \includegraphics[width=0.49\textwidth]{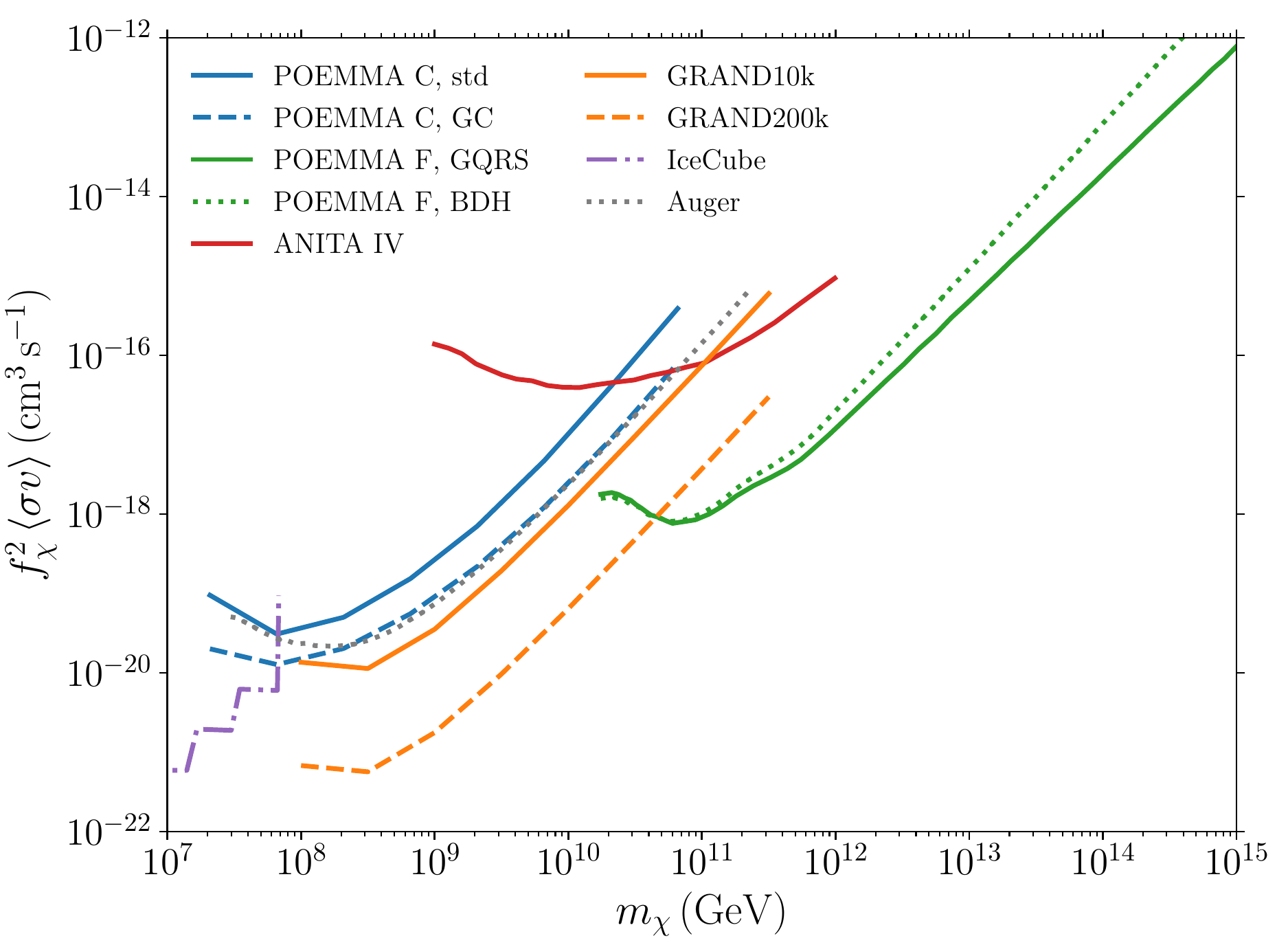}
    \includegraphics[width=0.49\textwidth]{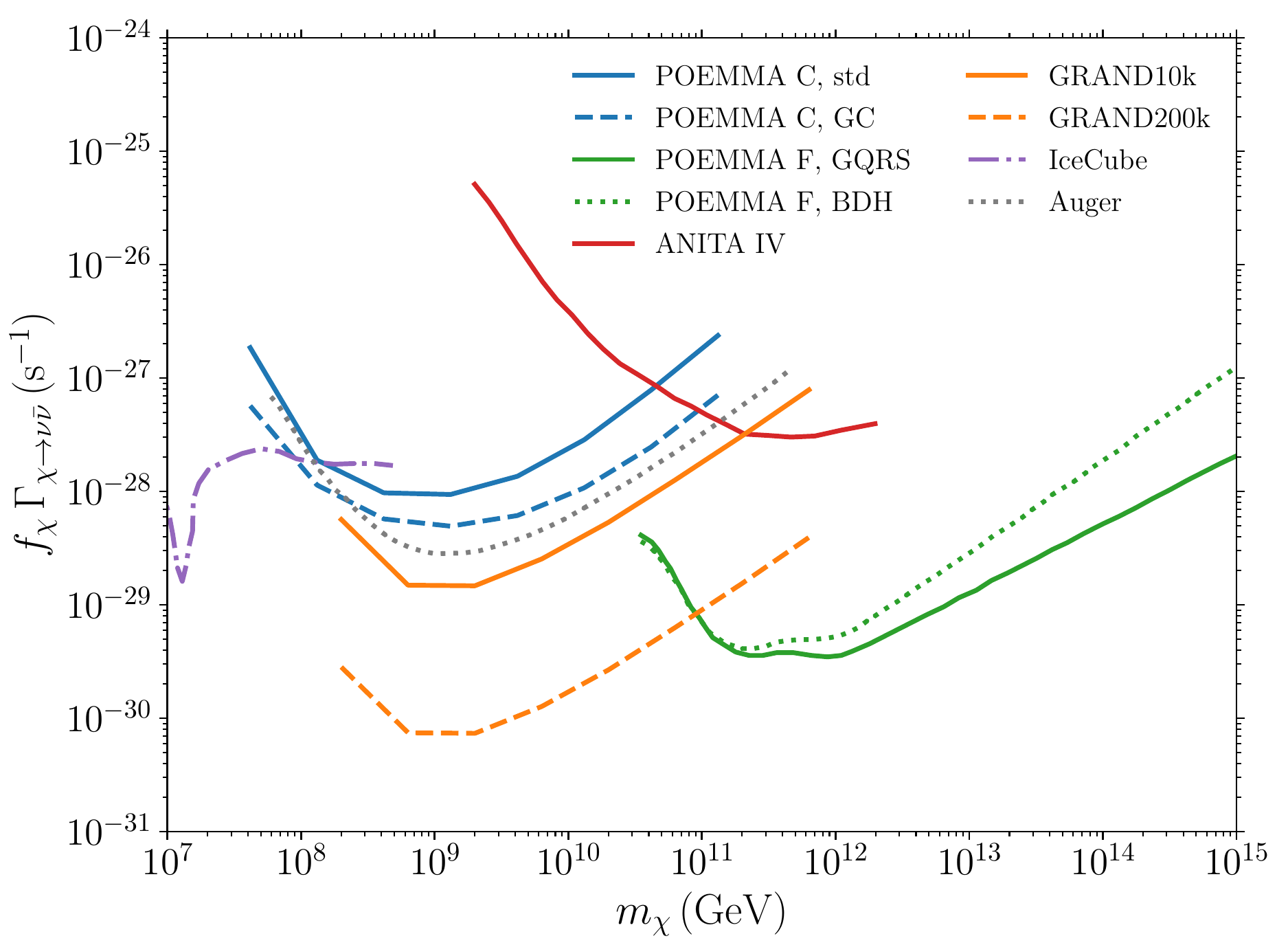}
    \caption{Sensitivities to dark matter thermally averaged annihilation cross section (left) and to dark matter decay width (right) for the $\nu\bar{\nu}$ channels. We show the $5$-year sensitivities of POEMMA for the Cherenkov standard (std, solid blue line), Cherenkov Galactic center (GC, dashed blue line), and the fluorescence (green lines) observation modes, GRAND10k (solid orange line) and GRAND200k (dashed orange line). We also show the sensitivity of ANITA IV (red line), Auger (dotted gray line), and IceCube (dot-dashed purple line).}\label{fig:CS_obs}
\end{figure}

The $5$-year sensitivities for POEMMA and GRAND, and the updated limits for Auger and ANITA, for the dark matter annihilation cross section and the decay width, are illustrated in figure~\ref{fig:CS_obs}. Overlayed are the IceCube limits, respectively \cite{Arguelles19} for annihilation, and \citep{IceCube18} for decay. These limits are corrected to account for the different dark matter distribution used, using the ratios between the line-of-sight integrals of the dark matter profiles.

Our results show that GRAND (GRAND200k) will achieve the most constraining bounds in the energy range $\sim 10^8-10^{11}\,{\rm GeV}$, due to its large exposure, detection from all azimuth angles and a full time operation. The location of the antennas arrays, the efficiency of autonomous radio detection and level of noise, to be tested with the next phase GRANDProto300 \citep{Decoene19ICRC}, will determine the ability of GRAND to improve the existing limits by two orders of magnitude.

POEMMA has the advantage of full-sky coverage and pointing ability in the Cherenkov detection mode, allowing to develop optimized observation strategies. The Galactic center observation mode can increase POEMMA's sensitivity, improving by a factor $\sim 2$ at $10^9\,{\rm GeV}$ the constraint derived by the IceCube Collaboration \cite{IceCube18} for dark matter decay. The Galactic center observation mode is particularly advantageous for a dark matter distribution peaked towards the Galactic center. Above $10^{11}\,{\rm GeV}$, POEMMA's fluorescence observation mode has an unprecedented sensitivity to superheavy dark matter properties, improving by a factor of $\sim 80$ the sensitivity of ANITA-IV.

We compare our limits for Auger with existing constraints, from \cite{Arguelles19} for the annihilation channel, and \cite{Esmaili12, Kachelriess18} for the decay channel. After including numerical factors accounting for the different dark matter profiles and the increase of exposure with time, we find large discrepancies with \cite{Arguelles19} (annihilation) and \cite{Esmaili12} (decay), which could be related to a discrepancy in the calculation of the time-average sky coverage of the detector. Our calculation differs from \cite{Kachelriess18} (decay) by only a factor of $2$ in the mass range $10^8-10^{11}\,{\rm GeV}$, and does not include a high energy tail above $10^{11}\,{\rm GeV}$ related to the distribution of secondary neutrinos.

\section{Sources of uncertainties}

\begin{figure}[ht]
    \centering
    \includegraphics[width=0.49\textwidth]{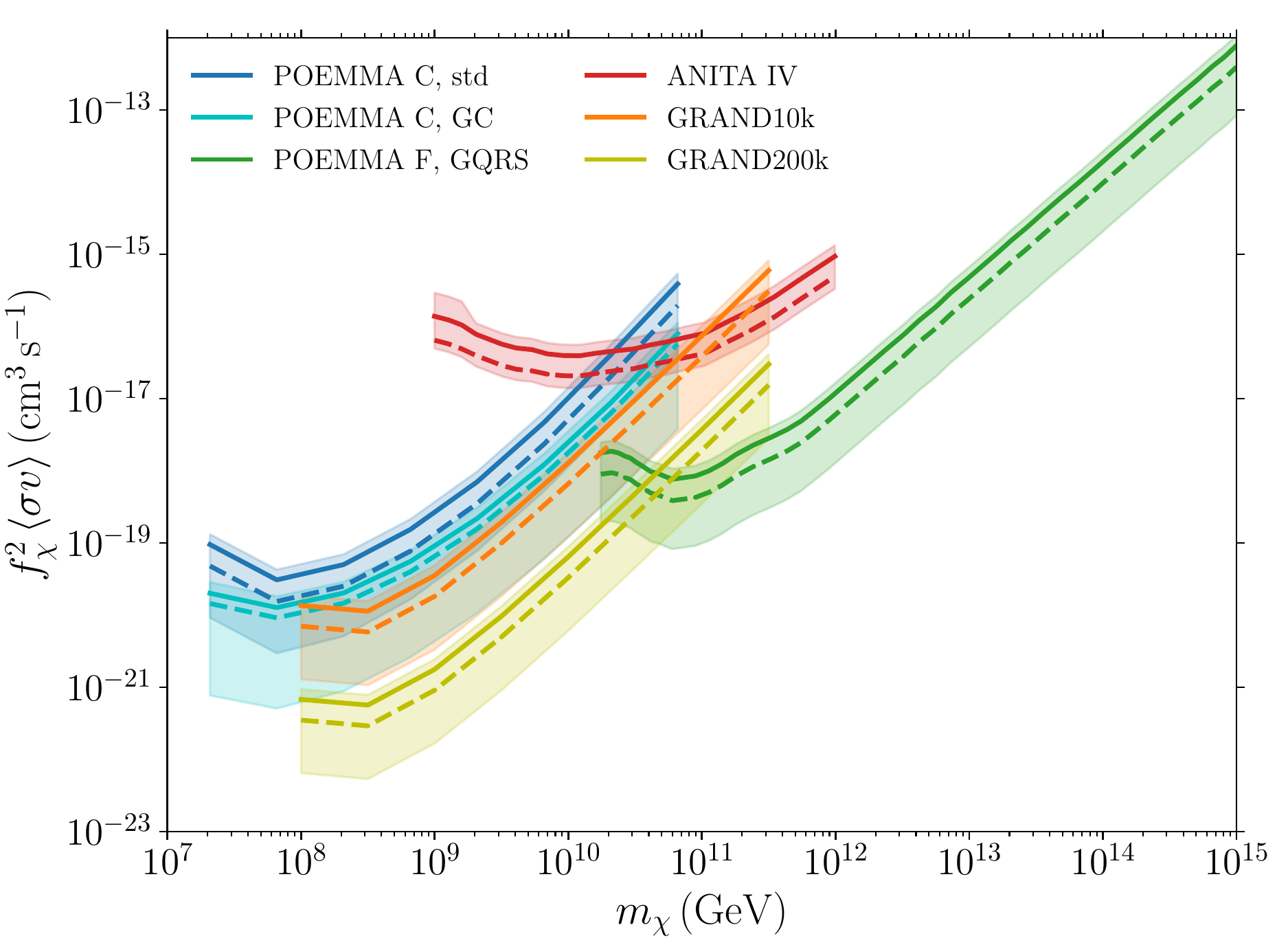}
    \includegraphics[width=0.49\textwidth]{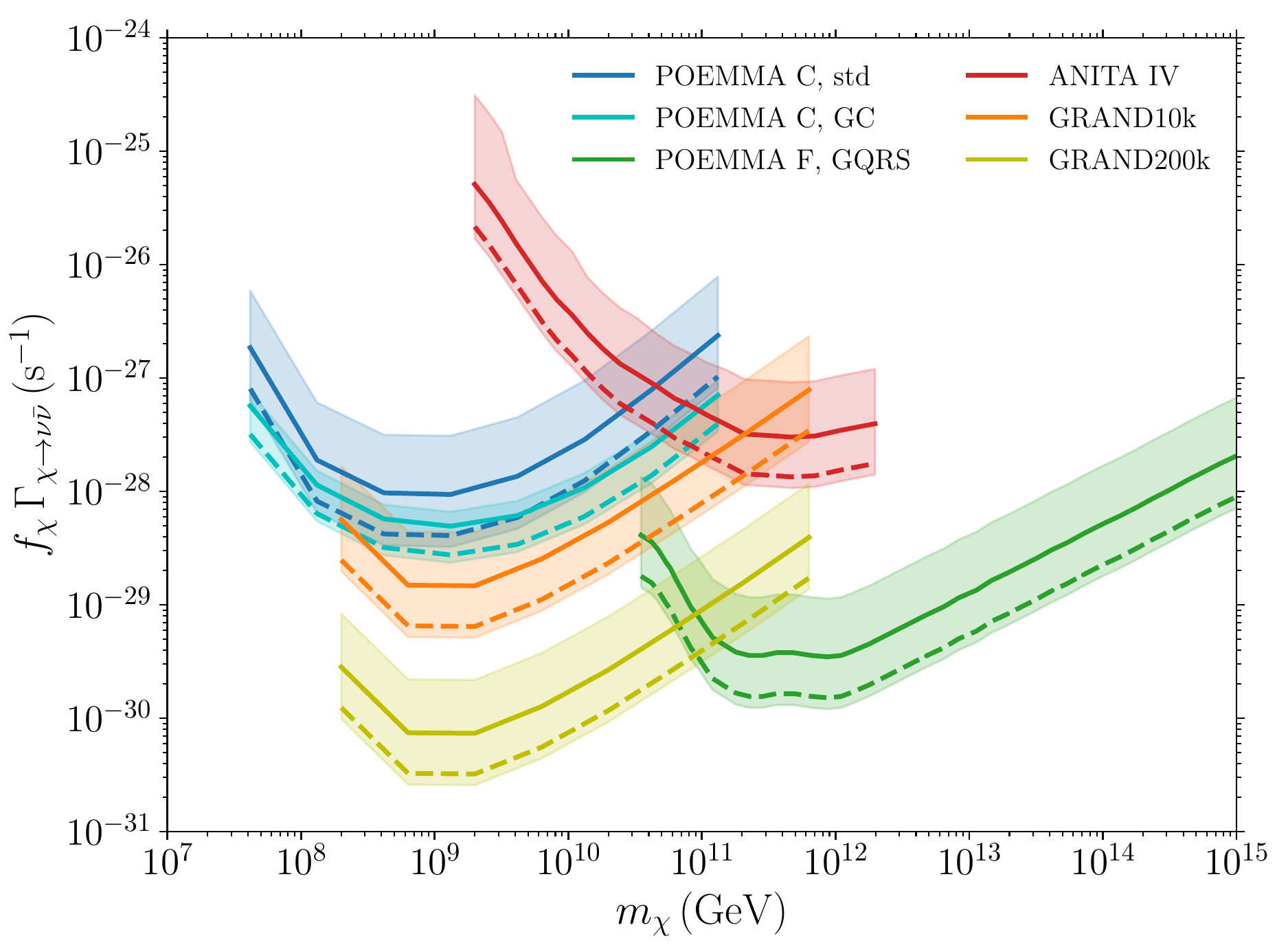}
    \caption{Uncertainties on the sensitivities to dark matter thermally averaged annihilation cross section (left) and on the sensitivities to dark matter decay width (right), for the $\nu\bar{\nu}$ channels.}\label{fig:uncertainties}
\end{figure}

The limits presented in figure~\ref{fig:CS_obs} are calculated using the Burkert dark matter profile. However, the dark matter profile is loosely constrained by rotation curve measurements, thus we calculate the uncertainties related to this profile. These uncertainties have been estimated in several previous studies: for instance \cite{IceCube18} obtain $\pm 10 \%$ uncertainties on the dark matter lifetime, and \cite{Arguelles19} find $\sim 1$ order of magnitude uncertainties on the dark matter annihilation cross section. In this work, we calculate the $1\sigma$ uncertainties associated with the tabulated parameters $\rho_0$, $\gamma$, $R_s$ and $V_0$ (the circular velocity of the Sun) of the generalized NFW distribution, using the general fit presented in \cite{Benito20}, such that $\chi^2 - \chi^2_{\rm best \, fit} < 4.72$. These uncertainties are represented by the bands in figure~\ref{fig:uncertainties}, while the dashed lines are the sensitivities obtained for the best fit parameters from \cite{Benito20} and the solid lines are the sensitivities obtained with the Burkert distribution. We find $1\sigma$ uncertainties of $1-1.5$ orders of magnitude, with variations depending on the sky coverage of the detectors: for instance the POEMMA Cherenkov Galactic center observation mode is only sensitive to the dark matter distribution around the Galactic center, and thus uncertainties for this observation mode are larger for the annihilation channel due to the factor $\rho_\chi^2$ in Equation~\ref{eq:num_tau}.

We evaluate the uncertainties related to the distribution of secondary products, which can be influenced by electroweak showers by the introduction of neutrinos at lower energies. We use recent calculations including electroweak fragmentation function evolution, matching at the weak scale, and evolution with Pythia \citep{Bauer20}. We find a small difference between the delta function approximation and the distribution function including cascades, compared to uncertainties related to the dark matter profile. The main difference is a production of high energy tails in the sensitivities, related to the low-energy tail of the secondary neutrino distribution. However, these high energy tails produce low sensitivities in the mass range considered, and can thus be neglected.

\section{Motivation for future experiments}

\begin{figure}[t]
    \centering
    \includegraphics[width=0.30\textwidth]{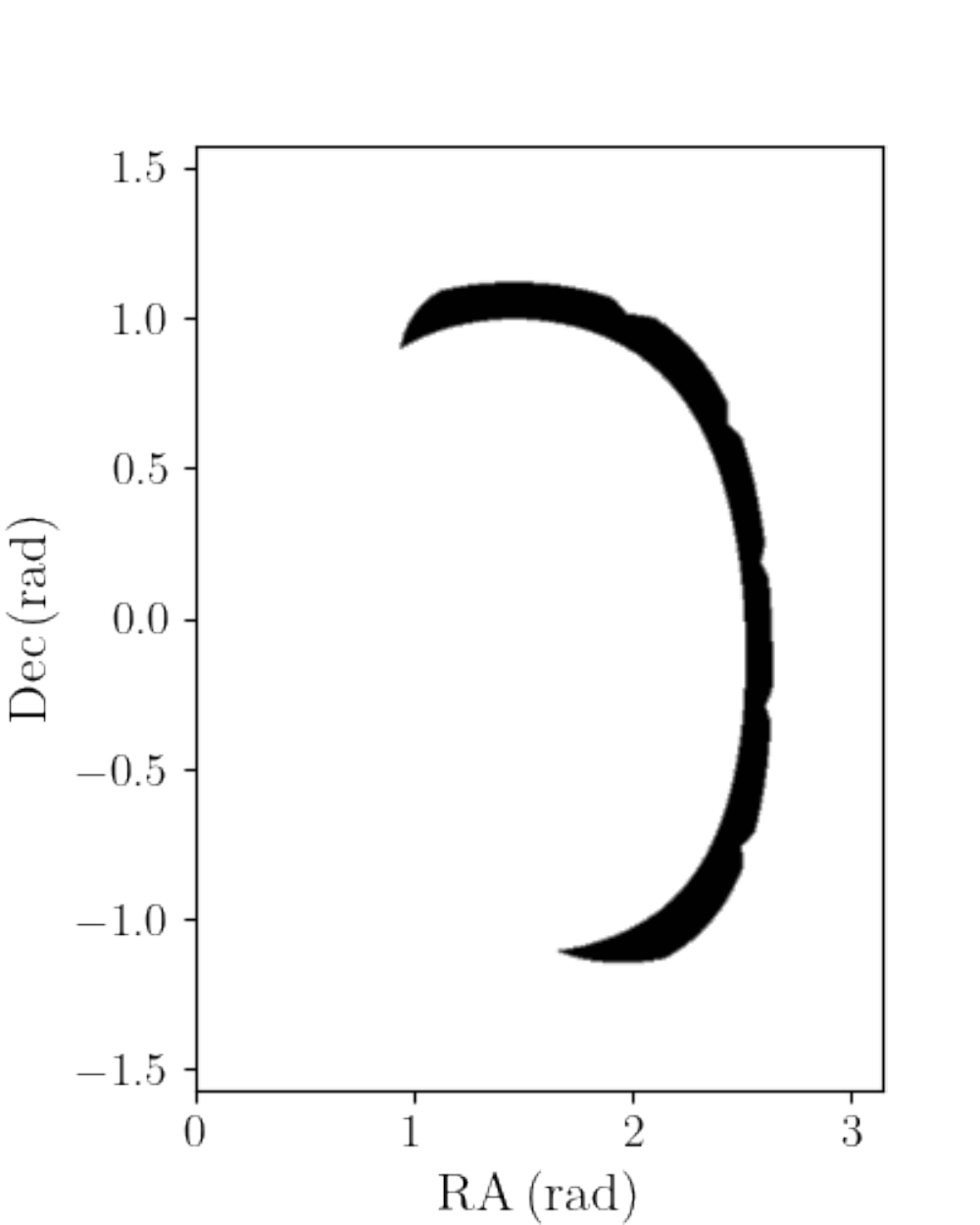}
    \includegraphics[width=0.6\textwidth]{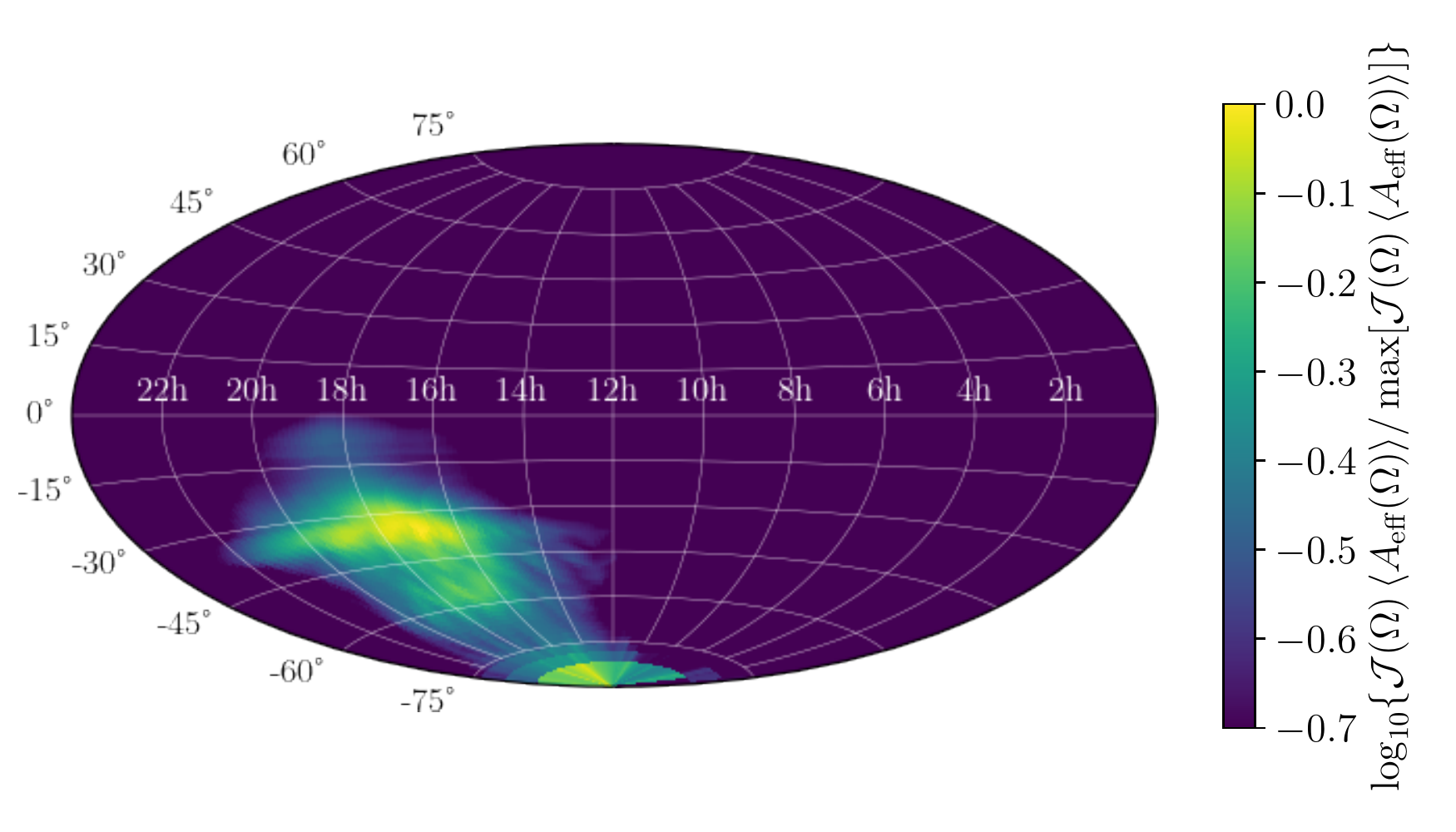}
    \caption{Illustration of the configuration with six POEMMA-like Cherenkov detectors: instantaneous field of view as a function of right ascension (RA) and declination (DEC) (left) and effective area weighted by ${\cal J} = \int_{\rm l.o.s.} {\rm d}x \, \rho_\chi^2 (x)$ and normalized, for $E_\nu = 10^{8.5}\,{\rm GeV}$ (right).}\label{fig:sky_cov_detectors}
\end{figure}

Various projects of HE-UHE neutrino detectors are planned for the next decades, such as IceCube-Gen2 \citep{IceCubeGen2}, RNO-G, \citep{Aguilar21RNOG}, Trinity \citep{Otte19}, and others \citep{Neronov19}. These neutrino detectors will profitably contribute to dark matter searches in the $m_\chi = 10^8-10^{15}\,{\rm eV}$ mass range. In this context, we also examined how a detector comprised of several POEMMA-like Cherenkov detectors pointing in different directions (in order to increase the field of view along the limb) can increase the sensitivity to superheavy dark matter properties. The case of six detectors is illustrated in figure~\ref{fig:sky_cov_detectors}. We show that the uncertainties on the dark matter profile strongly influence the estimates of the sensitivity gains, with smaller gains for peaked dark matter distribution towards the Galactic center. For the case of six POEMMA-like detectors, covering $\sim 180\degree$ in azimuth, using the best fit parameters of the generalized NFW distribution \citep{Benito20}, we find an enhancement in the sensitivity by a factor $4$ for annihilation and $6$ for decay.

\vspace{0.5cm}

{\bf Acknowledgements}

The authors thank Francis Halzen, Cosmin Deaconu and María Benito for useful discussions. C.G. is supported by the Neil Gehrels Prize Postdoctoral Fellowship. L.A.A. is supported by the U.S. National Science Foundation (NSF) Grant PHY-2112527. M.H.R. is supported in part by U.S. Department of Energy Grant DE-SC-0010113.

\bibliographystyle{apsrev}
\bibliography{SHDM}

\end{document}